\documentstyle[11pt,epsf,fullpage]{article}

\title{{\bf Pion Lattices in HI Transverse Phase Space and Sudden Traversal of the QCD Phase Boundary}}

\author{{\bf Thomas A. Trainor}\\ \\
{\em Nuclear Physics Laboratory 354290}\\
{\em University of Washington}\\
{\em Seattle, WA 98195}\\
{\em trainor@hausdorf.npl.washington.edu}}
			       
\def\jnl#1#2#3#4{{#1} {\bf #2}, #3 (#4)}

\def\npa{{ Nucl. Phys.} A}
\def\plb{{ Phys. Lett.} B}
\def\prl{ Phys. Rev. Lett.}

\def\be{\begin{equation}}
\def\ee{\end{equation}}
\def\bea{\begin{eqnarray}}
\def\eea{\end{eqnarray}}
\newcommand{\xfig}[1]{\begin{center}
\mbox{\epsffile{#1}}
\end{center}}

\begin{document}
\maketitle

\begin{abstract}
Correlation data from high-energy p-p collisions indicate substantial anticorrelation of like-sign pions and correlation of unlike-sign pions on rapidity and azimuth at small length scale suggesting approach to a 1D lattice structure during string hadronization. I generalize this result to propose that an {\em incipient} 3D pion lattice structure may be formed during hadronization of a QGP. I then present a method to detect this structure in the transverse phase space of heavy-ion collisions which depends on local transverse Hubble flow. I also relate this correlation structure to recently proposed suppression of net charge and baryon-number fluctuations on rapidity after sudden traversal of the QCD phase boundary.
\end{abstract}

%\section{Introduction}

The search for definitive signatures of QGP formation in heavy-ion collisions based on event-by-event fluctuation analysis is intensifying.
%\cite{list}. 
Attention has focused on critical fluctuations (an excess) expected near the QCD phase boundary, and especially near a proposed critical end point \cite{mish}.  Recent predictions of {\em reduced} multiplicity fluctuations as a QGP signal emphasize a different aspect of fluctuation phenomena related to the small-scale structure of the hadronic final state. In this letter I discuss fluctuation analysis as initially applied to critical fluctuations, consider recent predictions of fluctuation suppression associated with QGP formation, describe this suppression in terms of an incipient pion lattice in configuration space, comment on the observability of suppression on rapidity, relate pion lattice formation to correlation structures in p-p collisions and propose that an incipient pion lattice would best be observed in combination with nuclear Hubble flow as a quadratic correlation structure in {\em transverse} two-particle phase space. Such structure would serve as a robust indicator for an extended prehadronic state in heavy-ion collisions.

%\section{Fluctuation analysis}

Fluctuation analysis has previously addressed the question of equilibration and critical fluctuations in HI collisions. A generalized central limit theorem is the primary fluctuation reference for a scale-invariant equilibrium hypothesis \cite{tatclt}. Two recent papers \cite{asa,koch} focus on net fluctuations of bipolar, locally-conserved dynamical measures on rapidity as possible indicators of a deconfined partonic system (QGP) produced in heavy-ion (HI) collisions. The authors state that such fluctuations reflect the microscopic structure of nuclear matter and argue that because the density of dynamical degrees of freedom (DoF) in a QGP is significantly larger than in a hadron gas (HG) measure fluctuations in the QGP phase should be suppressed relative to the hadronic phase for equivalent systems. They argue that if the transition from QGP to decoupled hadrons is sufficiently  rapid a significant fraction of this suppression may survive to the final state, providing a `clear signal' for  QGP production.  In \cite{asa} net baryon-number fluctuations in  a QGP were determined to be about 2.4 times less and net charge fluctuations about 1.6 times less than in a HG based on  a parton counting argument (uncorrelated parton gas). In \cite{koch} QGP fluctuation estimates were based on an entropy argument assuming an ideal parton gas and also derived from lattice calculations.   The lattice result at finite temperature gave less reduction (1/4) than the IG estimate (1/5), due to quark correlations. Summarizing predictions: fluctuations of net charge and net baryon-number multiplicities in a rapidity interval of 1-2 units should be suppressed by factors of $2 - 5$ relative to an ideal HG in a hadron system produced by sudden traversal of the QCD phase boundary  according to these authors.   

\begin{figure}[th]
\begin{tabular}{cc}
\begin{minipage}{.57\linewidth}
\epsfysize .5\textwidth
\xfig{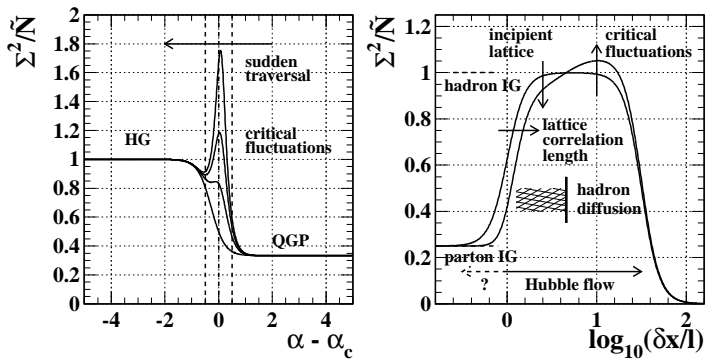}  
%\caption{\label{}}
\end{minipage} &
\begin{minipage}{.37\linewidth}
\epsfysize 1\textwidth
%\pawplot{.ps}  
\caption{Total variance {\em vs} constraint $\alpha$ near the QCD phase boundary (left) showing reduced variance for increased DoF in the QGP phase. Large-scale correlations at the phase boundary result in `critical' fluctuations. Scale dependence of total variance (right) indicates possible reduction of variance from a HG value by lattice formation depending on lattice correlation length. \label{boundary}}
\end{minipage}
\end{tabular}
%\caption{\label{}}
\end{figure}

These arguments are illustrated in the left panel of  Fig. \ref{boundary}: a step decrease in total variance $\Sigma^2$ \cite{tatclt} from HG to QGP regions of a system constraint $\alpha$ and critical fluctuations at the phase boundary. Any large-scale critical correlations developed at the phase boundary reduce the effective number of DoF and increase fluctuations. We ask whether the system remembers a prehadronic correlation state (vertical dashed lines) after sudden traversal of the phase boundary \cite{raja}.  Both increased and decreased variance are possible depending on traversal rate and observation scale, making a scale-dependent fluctuation analysis essential \cite{tatclt,tatsomething}. The argument in \cite{asa}  for observability of these reductions in the final state is based on competition between Bjorken (longitudinal Hubble) expansion and thermal motion (diffusion).  Attenuation of the effect by rescattering is said to be reduced by observing fluctuations in a rapidity interval `much larger' than unity.
%%%%%%%%%%%%%%%%%%%%%%%%%%%%%%%
%%%%%%%%%%%%%%%%%%%%%%%%%%%%%%%

%\section{Fluctuations and Correlations}

Fluctuations describe the correlation structure of a measure distribution binned at a specific scale (bin size).  For a nearly equilibrated composite system of unipolar and bipolar measures fluctuations are described in a gaussian model by a scale-dependent covariance matrix \cite{tatclt}.  {\em Specific} measures $\hat m \equiv m/\bar m$ are reported relative to their mean values, with $\delta \hat m \approx \delta log(\hat m)$,  $\delta (\hat m_1 - \hat m_2) \approx \delta (\hat m_1 / \hat m_2)$ and  $\delta (\hat m_1 + \hat m_2) \approx \delta (\hat m_1 \cdot  \hat m_2)$. Small-amplitude fluctuations in sums and differences are equivalent respectively to fluctuations in products and ratios. A substantial analysis would consider fluctuations in each individual measure and in all sum/products and difference/ratios of measures taken in pairs.  For the ideal-gas systems considered in \cite{asa,koch} fluctuations in differences/ratios and sums/products are equivalent, implying that fluctuations in {\em all}  measure combinations should be comparably suppressed. Thermal meson resonances in a HG by contrast suppress ratio/difference fluctuations but increase product/sum fluctuations \cite{mish,koch2}. We must consider {\em both}  sum/product {\em and} difference/ratio  fluctuations to distinguish correlation sources.

 {\em Suppression} of hadronic net-charge and net-baryon-number fluctuations (difference/ratio fluctuations) relative to an ideal HG implies  reduced correlation compared to a randomly distributed (uncorrelated) system, equivalently an {\em anticorrelated} distribution. It follows immediately that such a distribution must approach a lattice configuration as the limiting case. I now extend the proposed argument:  A fundamental indicator of QCD phase-boundary penetration should then be {\em local asymptotic approach to a lattice configuration} in the configuration-space distribution of final-state hadrons. The right panel in Fig. \ref{boundary} illustrates the consequences of an incipient pion lattice. Hadronic fluctuations are suppressed toward a partonic value from the hadron separation scale $l$ up to some lattice correlation length, an {\em upper} scale limit  depending on traversal time and combinatoric issues which is unlikely to be large. Observational access on rapidity is limited by thermal diffusion which imposes an effective {\em lower} scale limit. It is possible that conflict of limits may preclude the observability of fluctuation suppression in rapidity bins. I argue below that an alternative manifestation of incipient lattice formation $via$ local Hubble flow can provide access down to the $fm$ length scale independent of diffusion and rescattering.

 The proposed suppression mechanism is equivalent to Kadanoff block scaling of spins: more numerous partons coalesced to fewer hadrons result in suppression of hadronic position fluctuations (anticorrelation) relative to random as a relic of the greater position symmetry of the precursor system (local entropy conservation). This implicitly models hadronization as instantaneous uniform spatial partitioning of  a deconfined parton fluid as opposed to less likely coalescence of point objects, but does not provide a mechanism for partition uniformity. It serves then as a limiting case. As an alternative correlation mechanism I consider the equilibrated QGP as an isoscalar with isovector and isotensor fluctuations. In a sudden traversal of the QCD phase boundary isospin fluctuations in the emerging pion fluid may mimic the greater symmetry of the prehadronic state by increased anticorrelation of like-sign pairs and correlation of  unlike-sign pairs. These (anti)correlations should be most significant at the hadronic length scale - 1 $fm$. Excess isotensor fluctuations have been the subject of recent DCC searches as a manifestation of chiral {\em critical} fluctuations. Suppressed isovector fluctuations (fluctuations in net pion charge) may as well reflect a chirally symmetric precursor system at small scale after sudden traversal of the QCD phase boundary, with the possibility of asymptotic approach to an {\em isospin antiferromagnet}: two charge species form overlapping  like-sign arrays locally approximating lattices, with unlike-sign nearest neighbors (neutral pions could be accommodated in this picture by considering local isotensor fluctuations).  The same argument may apply to other bipolar measures ({\em e.g.,} produced baryons/antibaryons, produced singly-strange/antistrange hadrons). This example extends the proposals of reduced fluctuations by presenting a detailed correlation/anticorrelation scheme related specifically to isospin symmetry. A confirmation of this picture for charged pions is found in p-p data.

%\section{p-p Correlations}

In \cite{pp} strong correlation structures were observed in the two-particle momentum density $\rho_2(y_1,y_2)$. Data corrected for event multiplicity fluctuations exhibit strong local structure on rapidity difference and azimuth-angle difference, with a correlation length of about 1 unit. Particle pairs with no {\em net} charge or transverse momentum (antilinear emission of unlike-sign hadrons) are apparently most favored, and pairs with net charge and transverse momentum (collinear emission of like-sign hadrons) are most hindered. The degree of (anti)correlation increases nonlinearly with event multiplicity, suggesting a strong density dependence.  Hadronic fluctuations immediately associated with a partonic precursor thus favor the most symmetric local phase-space configurations of dynamical measures, consistent with a statistical model applied to hadronic abundance distributions for a wide range of collision systems, including p-p collisions \cite{becc}. The common issue is hadron production from a locally equilibrated (maximally symmetric) prehadronic system. For a 1D color-deconfined string we therefore have evidence in hand for small-scale (anti)correlations in configuration space. We can extend these results to form an expectation about correlation structure arising from hadronization of 2D and 3D deconfined systems.

%\section{Phase-space Autocorrelations}

\begin{figure}[th]
\begin{tabular}{cc}
\begin{minipage}{.57\linewidth}
\epsfysize .5\textwidth
\xfig{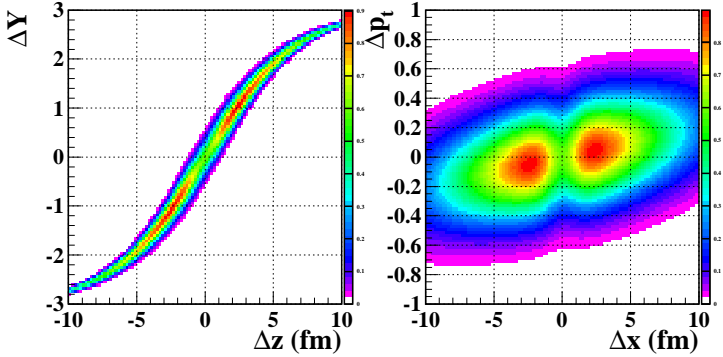}  
%\caption{\label{}}
\end{minipage} &
\begin{minipage}{.37\linewidth}
\epsfysize 1\textwidth
%\pawplot{.ps}  
\caption{Like-sign phase-space autocorrelation densities $g_-({\bf y,q})$ at kinetic freezeout for longitudinal p-p (left) and transverse HI (right) phase spaces. For the p-p case like-sign anticorrelation is resolved on rapidity. For the HI case anticorrelation is not resolved on $p_t$ directly, but a quadratic distortion is produced $via$ the $cosh$ factor in Eq. (\ref{threefac}).\label{g-qy}}
\end{minipage}
\end{tabular}
%\caption{\label{}}
\end{figure}

Direct observation of possible configuration-space correlation structures is reasonably attainable for p-p collisions, as illustrated in the left panel of Fig. \ref{g-qy}. This figure represents autocorrelation densities at freezeout for p-p axial phase space (left) and HI transverse phase space (right).  Because a high-energy p-p collision forms a 1D string with strong axial Hubble flow we have  direct access to small-scale configuration-space correlations {\em via} the rapidity autocorrelation, but this is not the case for HI collisions. A direct observation scheme applied to rapidity in HI collisions has difficulties beyond the issue of hadronic diffusion:  combinatoric attenuation of axial correlation structures caused by large-scale lattice deviations and/or uncorrelated string superposition resulting from projection of a 3D lattice to 1D.  Superposing 100 random strings with 10 hadrons per string gives 9k correlated {\em sibling} pairs but 1 M  random combinatoric pairs: a  S/N ratio of 1/100. For an extended QGP in a substantial 3D volume the situation could be more favorable, but large-scale lattice errors could also produce a nearly random hadronic distribution at small scale in the 1D projection.  These are daunting  obstacles to direct observation of lattice correlation effects or reduced fluctuations on rapidity.  

The central issue in HI collisions is the correlation structure not of longitudinal but of  {\em transverse} phase space. Are strings cross coupled, is color transversely deconfined over much of the collision system \cite{paj}? Evidence for small-scale (anti)correlation in HI collisions would be therefore most significant in the transverse system. We require  a means to observe small-scale structures in transverse phase space which probes directly the local  3D structure of  hadronic correlations. I now consider local 3D  Hubble flow observed in HI collisions.
 %%%%%%%%%%%%%%%%%%%%%%%%%%%%%%%
%%%%%%%%%%%%%%%%%%%%%%%%%%%%%%%
%\section{Hubble Flow}
We can search for incipient lattice formation in HI collisions in two ways:  1) indirectly by detecting reduced multiplicity fluctuations in a rapidity bin and 2) directly {\em via} a manifestation of local 3D Hubble flow in two-particle momentum autocorrelation densities. 
The right panel of Fig. \ref{g-qy} represents the transverse phase space of a central HI collision  where Hubble flow is subthermal. Local configuration-space  anticorrelation (distribution minima at $\Delta z$, $\Delta x = 0$) is not directly observable in momentum space. However, information on small-scale configuration-space correlation {\em is} accessible $via$ Hubble flow.   

The contribution of Hubble flow to the two-particle momentum distribution is contained in \cite{tathubble}
\begin{eqnarray} \label{threefac}
A({\bf k,q})  &\approx& m_{t1} \, m_{t2} \, e^{-{m_{t1} + m_{t2} \over T}}  \int \! d^4x g_+(x,k) \, exp\left( {2 H \over T }\, {{\bf k} \cdot {\bf x} } 
\right)  \int \! d^4y \,   g_-(y,q) \, cosh\left({{H \over 2 T}  \, {{\bf q} \cdot {\bf y} } } \right)
\end{eqnarray}
where $x = (x_1 + x_2)/2$, $y = y_1 - y_2$, are mean and relative four-positions, $k = (p_{1} + p_{2}) / 2$, $q = p_{1} - p_{2}$ are mean and relative four-momenta, $H$ is the local Hubble constant and $T$ is the local temperature. This is the {\em noninterference} term in a generalized form of the two-point momentum distribution $P_2({\bf p}_1,{\bf p}_2) = A + B$. The first two factors of $A$ represent the Boltzmann distribution on pair mean momentum and include the familiar radial-flow-induced blue shift. The third factor is newly formulated and represents a quadratic distortion of  the pair distribution on momentum difference which depends on Hubble flow {\em and} (anti)correlation in the sibling-pair autocorrelation. This factor contributes a positive or negative quadratic offset to the two-particle correlator on pair-momentum difference if {\em both} are present.

Comparing results for p-p and HI collision systems we can make the following observations concerning the third factor in Eq. (\ref{threefac}).  The autocorrelation density $g_-(y,q)$  can be treated in two limiting cases. For the p-p case we assume that axial Hubble flow is superthermal, as in the left panel of Fig. \ref{g-qy}. The autocorrelation density can then be approximated as $g_-(y,q) \rightarrow g(y) \, \delta(q - H\, y)$. In the HI case we assume that transverse Hubble flow is subthermal, as in the right panel of Fig. \ref{g-qy}, and that there is negligible momentum dependence: $g_-(y,q) \rightarrow g(y) $. In the first case spatial correlation structure in $g(y)$ is mapped directly to $P_2( \bar Y,\Delta Y)$ on rapidity. In the second case we obtain information on spatial correlation structure through a contribution quadratic in momentum difference from the $cosh$ factor whose amplitude then estimates the $rms$ width of $g(y)$ for sibling (same event) pairs relative to a mixed-pair reference.

Use of nuclear Hubble flow to resolve small-scale correlation structures on the hadronic freezeout surface is analogous to the similar effort in astronomy to map galactic positions {\em via} red shifts. It is possible that any quadratic bias built into the two-particle transverse momentum distribution by Hubble flow in combination with primordial particle correlations  may be quite robust against later hadronic rescattering, thereby providing access to local structures reflecting the hadronization process itself, and implicitly the existence of an equilibrated prehadronic state.

%\section{Conclusions}

In summary, I have considered recent predictions of rapidity fluctuations reduced  relative to a hadron ideal gas after sudden traversal of the QCD phase boundary. I interpret reduced fluctuations in terms of formation of an incipient pion lattice in configuration space but conclude that this phenomenon on rapidity may be greatly attenuated in HI collisions by longitudinal superposition of strings or by 3D lattice distortions as well as by thermal diffusion. Formation of an extended incipient pion lattice at the $fm$ scale {\em is} possible in high-energy heavy-ion collisions based on observed p-p rapidity (anti)correlations consistent with a maximum symmetry hypothesis. Although direct observation of corresponding structure on rapidity in HI collisions is doubtful, I have proposed an alternative observation mechanism in which nuclear Hubble flow combined with configuration-space (anti)correlations produces large-scale quadratic structures in two-particle momentum space. This mechanism may be robust against hadronic rescattering and may retain a substantial fraction of the local correlation information generated at hadronization.  The technique is applicable to transverse phase space and therefore extends the search for small-scale structure in a crucial way. Since {\em }rapidity (anti)correlations are already observed in p-p collisions their persistence  in HI collisions may not be remarkable. We require unique evidence in HI collisions for formation of a deconfined 3D partonic system. Detection of an incipient pion lattice in {\em transverse} phase space would offer conclusive indication of a prehadronic state in HI collisions.

\end{document}